\documentclass[epj]{svjour}

\usepackage{amsmath,amssymb}
\usepackage{array}
\usepackage{graphicx}
\usepackage{cite}

\renewcommand{\Im}{\mathop{\mathrm{Im}}}
\begin{document}

\title{Cross correlations in mesoscopic charge detection}

\author{Sigmund Kohler}
\institute{Instituto de Ciencia de Materiales de Madrid, CSIC,
Cantoblanco, E-28049 Madrid, Spain}

\abstract{
We study a tunnel contact that acts as charge detector for a
single-electron transistor (SET) focusing on correlations between the
detector current and the current through the SET.  This system can be
described fully by a Markovian master equation for the SET, while
electron tunneling in the charge monitor represents a process with a
stochastic rate, which can be solved exactly.  It turns out that current
monitoring is possible as long as the detector current correlates with
the currents through either SET barrier.  By contrast, correlations
with the effective current according to the Ramo-Shockley theorem are
not essential.  Moreover, we propose the measurement of the SET
barrier capacitances.
}

\date{\today}


\maketitle


\section{Introduction}

In experimental realizations of quantum dots, it has become standard
to place close to each dot a quantum point contact (QPC) for detecting
the charge state of the dots \cite{Gustavsson2006a, Fujisawa2006a,
Fricke2007a}.  Its working principle is based on Coulomb repulsion by
which the dot electrons reduce the conductivity of the QPC.  As an
alternative to a QPC, it has been suggested to employ a biased quantum
dot in which the interaction with a neighboring conductor may shift a
level across the Fermi surface \cite{Wiseman2001a, Schaller2010a}.
Also a double quantum dot may serve for this purpose if the proximity
of a charge detunes two resonant levels \cite{Kreisbeck2010a}.
Besides the direct measurement of charging diagrams, charge detectors
may be employed for quantum measurements in the transient regime such
as qubit readout \cite{Gurvitz1997a, Goan2001a, Wiseman2001a,
Gilad2006a, Ashhab2009b, Ashhab2009c, Kreisbeck2010a}, for testing
fluctuation theorems \cite{Sanchez2010a, Golubev2011a, Esposito2010a,
Campisi2011a} and dissipative effects \cite{Braggio2009a}, as well as
for inducing non equilibrium phenomena by direct energy transfer
\cite{Stark2010a, Hussein2012a} or via feedback loops
\cite{Schaller2011a}.

A quantum dot that is tunnel coupled to source and drain forms a
single-electron transistor (SET) through which electrons tunnel one by
one.  For weak coupling and certain bias voltages, this constitutes a
unidirectional stochastic process in which electrons enter the SET
exclusively from one lead while leaving to the other lead.  Then each
transition between the empty and the occupied SET can be attributed to
the tunneling of an electron either from the source to the SET or from
the SET to the drain.  Then the information provided by an attached
charge detector is sufficient to fully reconstruct the realization of
the underlying transport process.  In this way, one has determined
cumulants of the SET current \cite{Gustavsson2006a, Fricke2007a},
time-dependent full counting statistics \cite{Flindt2009a}, and
fluctuation spectra \cite{Ubbelohde2012a}.

\begin{figure}[b]
\centerline{\includegraphics{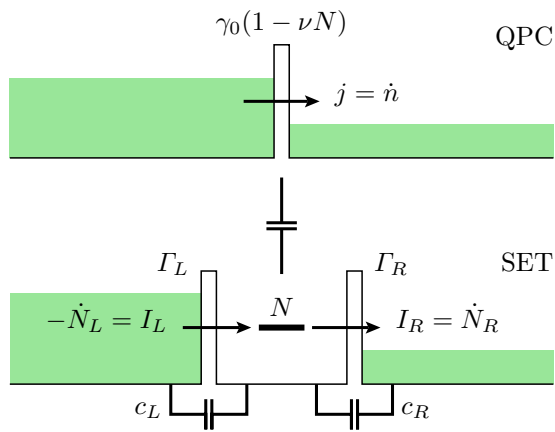}}
\caption{QPC in the tunnel regime acting as charge detector for a SET
with dot-lead tunnel rates $\Gamma_L$ and $\Gamma_R$.  The barrier
capacitances $c_L$ and $c_R$ determine the experimentally relevant SET
current $I = \alpha I_L+\beta I_R$, where $\alpha= c_R/(c_L+c_R)
=1-\beta$.  If the SET is empty, $N=0$, the number $n$ of electrons
transported through the QPC increases with the bare tunnel rate
$\gamma_0$.  Owing to the capacitive coupling between the two
subsystems, an electron on the SET reduces this rate by
$\nu\gamma_0$, where $\nu$ is the detector sensitivity.
}
\label{fig:model}
\end{figure}
Even though in these experiments, eventually the SET current is
determined, it is obvious that the measured quantity is the dot
occupation rather than the current.  Therefore, the question arises to
which extent the detector current correlates with the detected
current.  To this end, a SET with a QPC in the tunnel regime sketched
in Fig.~\ref{fig:model} represents a paradigmatic example, despite
that most experiments are performed in the limit of an open QPC.  A
main advantage of considering this model is the absence of backaction
and quantum mechanical superpositions, which facilitates the
interpretation.

In this work, we derive in Section~\ref{sec:model} a stochastic
process for the detector coupled to the SET.  In this model, the
detector does not act back to the SET occupation, so that we can
derive in Section~\ref{sec:SET} the correlation functions of the
latter before taking the detector into account.  The detector, by
contrast, is influenced by the SET and can be described by a Cox
process.  Its properties are derived in Section~\ref{sec:QPC}, while
Section~\ref{sec:cross} focuses on correlations between the
subsystems.

\section{SET coupled to a tunnel contact}
\label{sec:model}

The system sketched in Fig.~\ref{fig:model} consists of a SET formed
by a single-level quantum dot in contact with electron source and
drain.  Since double occupation of the SET is inhibited by Coulomb
repulsion, spin effects play a minor role and will be ignored.  This
setup is described by the Hamiltonian
\begin{equation}
\label{HSET}
H_\text{SET,leads} = \epsilon_0 N
+ \sum_{\ell,q} V_{\ell,q} (c^\dagger c_{\ell,q} + c_{\ell,q}^\dagger c)
+\sum_{\ell,q} \epsilon_q N_{\ell,q} ,
\end{equation}
where $c^\dagger$ and $c_{\ell,q}^\dagger$ are the usual fermionic
creation operators for an electron on the SET and in mode $q$ of
lead $\ell=L,R$, respectively, with the corresponding energies
$\epsilon_0$ and $\epsilon_q$ and the occupation numbers $N$ and
$N_{\ell,q}$.  Tunneling between the SET and the leads is
determined by the spectral densities $\Gamma_\ell(\epsilon) =
(2\pi/\hbar)\sum_q |V_{\ell,q}|^2 \delta(\epsilon-\epsilon_q) \equiv
\Gamma_\ell$ which we assume within a wideband limit energy
independent.

Within second-order perturbation theory in the SET-lead coupling
constants $V_{\ell,q}$, one can obtain a master equation for the
occupation probability $P_N$ of the SET, where $N=0,1$.  For
unidirectional transport, i.e., in the limit of large bias voltage, it
reads \cite{Gurvitz1996a}
\begin{equation}
\label{master}
\frac{d}{dt}\begin{pmatrix} P_0\\P_1 \end{pmatrix}
=
\begin{pmatrix} -\Gamma_L & \Gamma_R\\ \Gamma_L & -\Gamma_R \end{pmatrix}
\begin{pmatrix} P_0\\P_1 \end{pmatrix}
\equiv L P ,
\end{equation}
where the load rate $\Gamma_L$ and the unload rate $\Gamma_R$ are
given by the spectral densities of the respective dot-lead coupling.
The current through the left barrier is determined by the load rate
times the probability that the SET is empty, $I_L = \Gamma_LP_0$.
Vice versa, the current through the right barrier is given by $I_R =
\Gamma_R P_1$.  Notice that we consider particle currents, i.e, the
electrical current is obtained by multiplication with the elementary
charge.  

According to the Ramo-Shockley theorem \cite{Shockley1938a, Ramo1939a,
Blanter2000a, Mozyrsky2002a},
the experimentally measured current $I$ is the weighted average of the
currents through the left and the right tunnel barrier,
\begin{equation}
\label{RStheorem}
I = \alpha I_L + \beta I_R .
\end{equation}
The weights $\alpha$ and $\beta$ are determined by the barrier
capacitances between the SET and the leads and obey $\alpha+\beta=1$
\cite{Blanter2000a, Mozyrsky2002a}.  Typically, low-frequency
properties, such as the average current or the zero-frequency noise,
are the same for both $I_L$ and $I_R$.  Thus in that limit, one may
ignore the partition \eqref{RStheorem} or set for convenience both
weights to $1/2$.  Here however, we will find that some SET-QPC
correlations possess an imaginary part proportional to $\alpha-\beta$.

The charge detector is formed by a point contact in the weak coupling
limit which we model by a tunnel Hamiltonian whose transmission
depends on the SET occupation \cite{Gurvitz1997a, Braggio2009a}
\begin{equation}
\begin{split}
H_\text{QPC} = {}&
\sum_k \epsilon_k c_k^\dagger c_k
+\sum_{k'}\epsilon_{k'}c_{k'}^\dagger c_{k'}
\\&
+(1-\tilde\nu N)\sum_{kk'} T_{kk'} (c_k^\dagger c_{k'} + c_{k'}^\dagger c_k) .
\end{split}
\end{equation}
It couples the states $k$ of the left lead to the states $k'$ of the
right lead via the tunnel matrix element $T_{kk'}$, where
$c_k^\dagger$ and $c_{k'}^\dagger$ are the corresponding fermionic
creation operators for electrons with energies $\epsilon_k$ and
$\epsilon_{k'}$.  The number operator $N$ in the prefactor of the last
term reflects the fact that an electron on the SET reduces the tunnel
amplitudes.  The strength of this reduction depends on the interaction
with the SET which is quantified by the dimensionless parameter
$\tilde\nu$ .

Since $H_\text{QPC}$ commutes with $N$, the charge detector does not
directly act back to the SET occupation.  Notice that owing to
$[H_\text{QPC}, H_\text{SET,leads}] \neq 0$, there is an indirect
backaction which, however, is beyond second-order perturbation theory
in the SET-lead coupling.  Therefore it will not be considered.

In turn, within this level of description, the SET-lead tunneling
$V_{\ell,q}$ can be neglected in the computation of the QPC tunnel
rates.  Thus, we can adopt the golden-rule treatment of
Ref.~\cite{Ingold1992a} by which we obtain that an electron in state
$k$ of the left lead may tunnel to state $k'$ of the right lead with
probability $(2\pi/\hbar)|T_{kk'}|^2 \delta(\epsilon_k-\epsilon_{k'})
(1-\tilde\nu N)^2$.  Expressing the probability for the possible initial state
in terms of Fermi function and integrating over $\epsilon_k$ and
$\epsilon_{k'}$, we obtain for $N=0$ that the QPC current can be
described by a Poisson process \cite{Blanter2000a} with a rate
$\gamma_0$ proportional to the QPC bias voltage \cite{Ingold1992a}.
If an electron resides on the SET, i.e.\ for $N=1$, Coulomb repulsion
reduces the tunnel rates according to $\gamma_0 \to \gamma_1 \equiv
\gamma_0(1-\nu)$, where $\nu = \tilde\nu(2-\tilde\nu)$ reflects the
detector sensitivity and ideally assumes the value $\nu=1$.
Subsuming these two cases, we can conclude that
the QPC tunnel process inherits an additional randomness from
the SET occupation.  In more technical terms, the Poisson process
turns into a Cox process with a rate
\begin{equation}
\label{gamma(t)}
\gamma(t) = \gamma_0(1-\nu N(t))
\end{equation}
that depends on the two-state process $N(t)$.

\section{Charge-current correlations of the SET}
\label{sec:SET}

The SET described above is a frequently studied example of mesoscopic
transport for which most properties can be obtained analytically.  In
recent years, it has been employed for studying full-counting
statistics of the transported electrons, time-dependent correlations
\cite{Emary2007a}, as well as waiting time distributions \cite{Brandes2008a}.
Here by contrast, we are interested in the correlations between the
dot occupation and the incoming and the outgoing current.

Let us start by considering basic expectation values.  It is
straightforward to verify that the stationary solution of the master
equation \eqref{master} and the average currents read
\begin{equation}
\label{stationary}
P_0^\text{st} = \frac{\Gamma_R}{2\Gamma}, \quad
P_1^\text{st} = \frac{\Gamma_L}{2\Gamma}, \quad
\langle I\rangle = \frac{\Gamma_L\Gamma_R}{2\Gamma} 
= \langle I_L\rangle = \langle I_R\rangle,
\end{equation}
where $\Gamma \equiv (\Gamma_L+\Gamma_R)/2$.
An interesting observation is that in very asymmetric situations, $\Gamma
\approx \max(\Gamma_L,\Gamma_R)$, while $\langle I\rangle
\approx\min(\Gamma_L,\Gamma_R)$.

The master equation \eqref{master} provides direct information about the
SET only, while the lead degrees of freedom have been traced out.
Nevertheless, there exist various ways to obtain information about
the statistics of the transported electrons, e.g., by attributing a
counting variable to the terms that correspond to the tunnel process
of interest.  This provides the moment and the cumulant generating
function for the lead electrons in the long-time limit
\cite{Bagrets2003a} and for finite frequencies \cite{Emary2007a}.

Alternatively, one may employ the approach of Refs.\
\cite{Korotkov1994a,Hanke1995a}, which is rather convenient for
our purposes.  Its cornerstone is the conditional probability
$P(N,t|N',t')$ for having at time $t$ the SET occupation $N$ provided
that at an earlier time $t'$, it was occupied by $N'$ electrons.
Since the SET occupation is unique at any time, the conditional
probability must fulfill the boundary condition
$P(N,t'|N',t')=\delta_{NN'}$.  Moreover, it obeys the same master
equation as the unconditioned probability \cite{vanKampen1992a}.
Therefore, the conditional probability is equivalent to the propagator of
the master equation \eqref{master}, i.e.,
\begin{equation}
\label{conditional}
P(N,t|N',t') = \big[e^{L(t-t')}\big]_{NN'}\,,
\end{equation}
where
\begin{equation}
e^{Lt} = \frac{1}{2\Gamma}
\begin{pmatrix} \Gamma_R+\Gamma_L e^{-2\Gamma t} & &
                \Gamma_R & -\Gamma_R e^{-2\Gamma t} \\
                \Gamma_L-\Gamma_L e^{-2\Gamma t} & &
                \Gamma_L & +\Gamma_R e^{-2\Gamma t}
\end{pmatrix}
\end{equation}
can be computed readily by diagonalizing $L$.  Notice that the
conditional probability \eqref{conditional} is stationary, i.e., it
depends only on the time difference $t-t'$.

Before considering currents, we compute the auto correlation function
of the dot occupation, $\langle N(t)N(t')\rangle -\langle N\rangle^2$.
Since $N$ may assume only the values 0 and 1, the two-time expectation
value is given by the probability that the SET is occupied at both
times $t$ and $t'$.  Thus, in terms of the conditional probability and
the stationary occupation, it reads
$P(1,t|1,t')P_1^\text{st}$.  By use of Eqs.~\eqref{stationary} and
\eqref{conditional}, we find $C_{NN}(t-t') = (\langle I\rangle
/2\Gamma) \exp[-2\Gamma(t-t')]$.  Via Fourier transformation follows
the spectral density
\begin{equation}
\label{CNN}
C_{NN}(\omega) = \frac{2\langle I\rangle }{\omega^2+4\Gamma^2} \,.
\end{equation}

From the SET master equation \eqref{master}, we can draw conclusions
about the change of the lead electron number in the infinitesimal time
interval $[t,t+dt]$, $dN_R(t) = I_R(t)dt$.  This is possible because
owing to charge conservation, $dN_R(t) = -dN(t)$ for all transitions
that physically corresponds to electron tunneling between the SET and
the right lead.  Consequently, the expectation value $\langle
dN_R(t)\rangle$ is given by the joint probability that an electron
resides on the SET and that in the time interval $[t,t+dt]$, the SET
undergoes a transition from $N=1$ to $N=0$.  Thus, $\langle
dN_R(t)\rangle = P(0,t+dt|1,t) P_1^\text{st} = \langle I\rangle  dt$, with the
stationary current $\langle I\rangle $ given in Eq.~\eqref{stationary}.

The correlation between $N$ and $I_R$ can be obtained upon noticing
that the only trajectory contributing to the expectation value
$\langle N(t)\,dN_R(t')\rangle$ starts with
with an electron on the SET at time $t'$ which subsequently leaves to the right
lead during infinitesimal time $dt'$.  At time $t$, the SET must be occupied again.
The joint probability for this reads $P(0,t+dt|1,t) P(1,t|1,t')
P_1^\text{st}$.  Subtracting $\langle N\rangle\langle dN_R\rangle$ and
dividing by $dt$ yields the correlation function $C_{NI_R}(t-t') =
-\langle I\rangle P_0^\text{st} \exp[-2\Gamma(t-t')]$ for $t>t'$.  For $t<t'$, the
relevant trajectory starts with electron tunneling to the drain
followed by tunneling from the source.  This happens with
probability $P(1,t|0,t'+dt') P(0,t'+dt'|1,t')$, so that
\begin{equation}
\label{CNR}
C_{NI_R}(t) = \begin{cases}
+\langle I\rangle P_0^\text{st} e^{-2\Gamma|t|} & \text{for $t<0$}, \\
-\langle I\rangle P_1^\text{st} e^{-2\Gamma t} & \text{for $t>0$}.
\end{cases}
\end{equation}
By an analogous reasoning for the electron tunneling from the left
lead to the SET, we obtain
\begin{equation}
\label{CNL}
C_{NI_L}(t) = \begin{cases}
-\langle I\rangle P_1^\text{st} e^{-2\Gamma|t|} & \text{for $t<0$}, \\
+\langle I\rangle P_0^\text{st} e^{-2\Gamma t} & \text{for $t>0$}.
\end{cases}
\end{equation}
The anti-correlation between the current and the occupation manifest
in the negative sign of $C_{NI_R}(t)$ for $t>0$ has the obvious
interpretation that the SET is not occupied right after an electron
has tunneled to the right lead.  In the case of $C_{NI_L}(t)$, the
minus sign reflects the fact that an electron may tunnel to the SET
only when the latter is empty.  The by and large anti-symmetric
structure as function of time will be discussed below in the context
of SET-detector correlations.

Below we will need for the computation of normalized correlation
coefficients the auto correlation functions of the incoming and the
outgoing currents, $I_L$ and $I_R$.  They can be obtained from the
relation $dN_L(t)\,dN_L(t') = I_L(t)I_L(t')dt\,dt'$.  The only
contribution to this expression stems from two source-SET tunnel
events, one at time $t$, the other at $t'$. For $t>t'$ the
corresponding joint probability is $P(1,t+dt|0,t) P(0,t|1,t'+dt')
P(1,t'+dt'|0,t') P_0^\text{st}$.  Subtracting $\langle I\rangle^2$,
adding the shot noise $\langle I\rangle \delta(t-t')$,
and performing a Fourier transformation, we obtain the known result
\cite{Korotkov1994a, Blanter2000a, Emary2007a, Brandes2008a}
\begin{align}
S_{I_LI_L}(\omega)
={}& S_{I_RI_R}(\omega)
= \frac{\omega^2+\Gamma_L^2+\Gamma_R^2}{\omega^2+(\Gamma_L+\Gamma_R)^2}
  \, \langle I\rangle 
\\
\equiv {} & F_\text{SET}(\omega)\,\langle I\rangle  ,
\end{align}
where the frequency-dependent Fano factor is bounded according to $1/2
\leq F_\text{SET}(\omega) < 1$.  For for a derivation of the shot
noise term $\langle I\rangle \delta(t-t')$ in the spirit of the
present calculation, see Ref.~\cite{Korotkov1994a}.

\section{Detector current}
\label{sec:QPC}

Charge transport through the QPC in the tunnel limit is a Markov
process \cite{Blanter2000a}, where in our case the rate $\gamma(t) =
\gamma_0(1-\nu N)$ is a stochastic process as well, see Eq.~\eqref{gamma(t)}.
The state of the detector at time $t$ can be characterized by the
number of electrons $n$ that have been tunneling thus far.  For
unidirectional transport, a direct transition from $n$ is possible only to
$n+1$, but not to $n-1$.  One can readily write down a
master equation for the probability $p_n(t,t')$ that $n$ electrons
have been tunneling between time $t'$ and a later time $t$
\cite{vanKampen1992a},
\begin{equation}
\label{cox}
\frac{d}{dt}p_{n}(t,t')
= \gamma(t) \big[ p_{n-1}(t,t')-p_{n}(t,t')\big]
\end{equation}
with the initial condition $p_n(t',t') = \delta_{n,0}$.  For $\nu=0$,
$\gamma(t) = \text{const}$, so that this equation describes a Poisson
process, while otherwise, it constitutes a stochastic process
that inherits additional randomness from the SET.  It is
straightforward to demonstrate that the solution of the master equation
\eqref{cox} is the Poisson-like distribution
\begin{equation}
\label{pn}
p_n(t,t') = \frac{1}{n!}\phi_{t,t'}^n e^{-\phi_{t,t'}}
\end{equation}
with the mean number of events during time $t-t'$ replaced by the
integral
\begin{equation}
\label{phi}
\phi_{t,t'}
= \int_{t'}^t ds\,\gamma(s) \,.
\end{equation}

For the initial time $t'=0$, all moments and cumulants of this process
can be calculated in a more or less elementary way.  In particular,
one finds the expectation values
\begin{align}
\label{n}
\overline{n(t)} ={}& \phi_{t,0} ,
\\
\label{nn}
\overline{n(t)n(t')} ={}& \phi_{t,0}\phi_{t',0} + \phi_{\min(t,t'),0} .
\end{align}
The overbar denotes the average over only the tunnel contact, while
angular brackets will refer to the additional average over the full
system including the SET.

For completeness, we sketch the derivation of Eqs.~\eqref{n} and
\eqref{nn} and refer for a more general treatment to
Ref.~\cite{Bouzas2006a}.  The average number of transported electrons,
Eq.~\eqref{n}, is simply the first moment of the distribution function
\eqref{pn}.  The two-time expectation value \eqref{nn} can be defined
as $\sum_{n,n'} n n' p(n,t; n',t')$, where $p(n,t; n',t')$ is the
joint probability of finding $n'$ transported electrons at time $t'$
and $n$ electrons at a later time $t$.  With the help of Bayes
theorem, it can be written as $p(n,t; n',t') =
p(n,t|n',t')p_{n'}(t')$, where the conditional
probability $p(n,t|n',t')$ obeys the master equation \eqref{cox} with
the usual boundary condition $p(n,t|n',t) = \delta_{n,n'}$.  Thus, $p(n,t|n',t')
= p_{n-n'}(t,t')$, so that we find for $n\geq n'$ and $t\geq t'$ the
solution
\begin{equation}
\label{Pnn}
p(n,t;n',t') = \frac{\exp({-\phi_{t,0}})}{(n-n')!\, n'!}\, \phi_{t,t'}^{n-n'}
\phi_{t',0}^{n'}  \,.
\end{equation}
Evaluating the sum in the above definition yields Eq.~\eqref{nn}.

The detector current and its correlation function follow from
Eqs.~\eqref{n} and \eqref{nn} by time differentiation, which
yields expressions that still depend on the absolute time.  Only after
averaging over the SET variables, they become stationary and read
\begin{align}
\langle j\rangle ={}& \gamma_0(1-\nu P_1^\text{st}) , \\
C_{jj}(\omega) ={}& \langle j\rangle +\nu^2\gamma_0^2 C_{NN}(\omega) ,
\end{align}
where the latter is the Fourier representation of $C_{jj}(t-t')$.
An established measure for the relative noise strength is the
frequency-dependent Fano factor
\begin{equation}
\label{fano}
F_\text{QPC}(\omega)
= \frac{C_{jj}(\omega)}{\langle j\rangle}
= 1+\frac{\nu^2\gamma_0^2 C_{NN}(\omega)}{\langle j\rangle} .
\end{equation}
Its first term is the usual shot noise which one
obtains also for a constant QPC tunnel rate.  The second term is the
excess noise stemming from the stochastic character of QPC rate.
Generally, $F_\text{QPC}>1$ which indicates avalanche-like transport due to
the switching between high and low detector rates as a consequence of the
alternating SET occupation.
If the noise enhancement is small compared to the shot noise, the
detector current is essentially not affected by the SET, and we expect
that no information can be drawn from the measurement.  In accordance
with this, we will find that the correlation between the SET
occupation and the detector current is indeed governed by the Fano
factor \eqref{fano}.

\section{Detector-SET correlations}
\label{sec:cross}

From the expectation value \eqref{n} together with
Eqs.~\eqref{gamma(t)} and \eqref{phi} follows the detector current
$\overline{j(t)} = \gamma_0-\nu\gamma_0 N(t)$.  Since the first term of
this expression is constant, any correlation of $j$ with a SET
variable $x$ must stem from the second term and, thus, read
\begin{equation}
\label{Cjx}
C_{jx} = -\nu\gamma_0 C_{Nx}\,.
\end{equation}
Since we are interested in the degree of correlation rather than in
absolute values, we focus on the normalized correlation at a given
measurement frequency $\omega$ which we define as
\begin{equation}
\label{rjx}
r_{xy}(\omega) =
\frac{C_{xy}(\omega)}{\sqrt{C_{xx}(\omega)\,C_{yy}(\omega)}}\,.
\end{equation}
Its absolute value is a figure of merit for the detection quality and
in the ideal case is of order unity.  In turn, for $|r_{jx}|\ll1$, the
detector current is practically independent of $x$.

\subsection{SET charge}

By construction, the charge detector is sensitive to the SET charge,
which implies that the correlation coefficient $r_{jN}$ must be close
to unity at least in some frequency range.  Since the QPC current can
be expressed in terms of the SET occupation, we can express $C_{jj}$
and $C_{jN}$ by $C_{NN}$.  The latter can be eliminated in
favor of the frequency-dependent Fano factor \eqref{fano} so that
the correlation coefficient becomes
\begin{equation}
\label{rjN}
r_{jN}(\omega)
= -\sqrt{\frac{F_\text{QPC}(\omega)-1}{F_\text{QPC}(\omega)}} \,.
\end{equation}
It is close to unity if $F_\text{QPC}\gg1$.
This means that, as conjectured above, we can acquire information
about the SET occupation only if the detector current exhibits
significantly super-Poissonian noise.

\begin{figure}
\centerline{\includegraphics{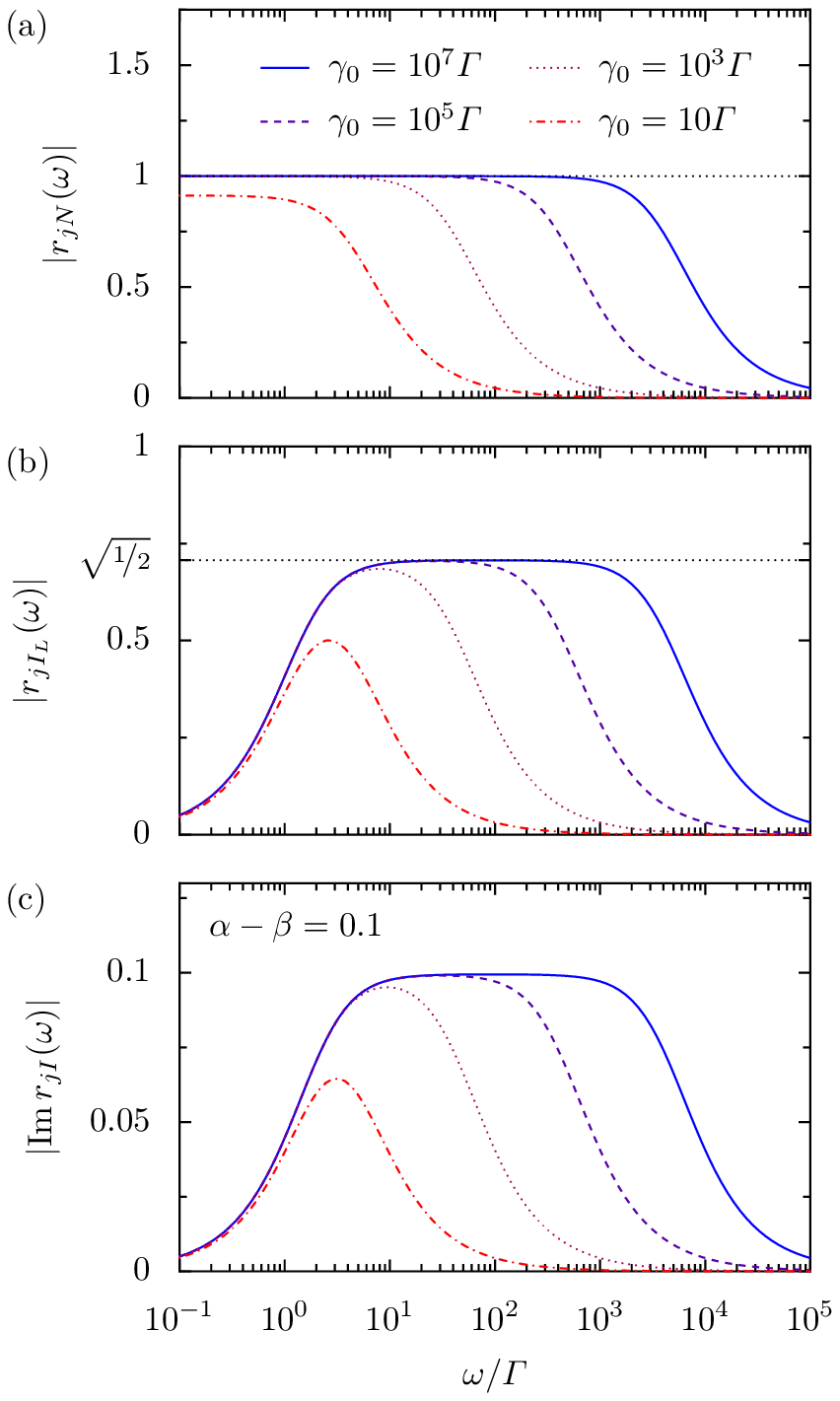}}
\caption{(a) Absolute value of the correlation coefficient between the
SET occupation and the detector current for various detector rates
$\gamma_0$, tunnel rates $\Gamma_L =\Gamma_R =\Gamma$, ideal
sensitivity, $\nu=1$, and $\alpha=\beta=1/2$.
(b) Absolute value of the correlation coefficient between the SET
source current $I_L$ and the detector current for the same parameters.
(c) Imaginary part of the correlation between detector current and the
total SET current for Ramo-Shockley coefficients with $|\alpha-\beta|
=0.1$.
}
\label{fig:correlations}
\end{figure}

Moreover, this result turns some natural expectations for the measurement
process into quantitative statements.  Fist, one expects that the time
resolution $\tau$ of the detector has some lower limit set by
the condition that during time $\tau$, many electrons should flow in
order to establish a clear signal.  Thus, $\tau\gg1/\gamma_0$ or in
terms of the measurement frequency: $\omega\ll\gamma_0$.  Second, for
the specific measurement of the SET occupation, the SET dwell time
should be larger than the time resolution, $1/\Gamma\gtrsim\tau$,
which together with the above condition yields $\Gamma\ll\gamma_0$.
The quantitative analysis shown in Fig.~\ref{fig:correlations}(a)
demonstrates that even for an ideal detector $\nu=1$, in order to find
$r_{jN}(\omega)\approx1$, these inequalities need to be fulfilled by
roughly two orders of magnitude, i.e., $\gamma_0\gtrsim 100\Gamma$ and
$\gamma_0\gtrsim100\omega$.

The Fano factor $F_\text{QPC}$ exhibits an interesting asymmetry with
respect to the SET tunnel rates.  For $\nu \approx 1$ and $\omega=0$,
it can be approximated as $F_\text{QPC} \approx
1+2\gamma_0\Gamma_L/\max(\Gamma_L,\Gamma_R)^2$.  Thus, starting from a
symmetric situation, the Fano factor remains constant if $\Gamma_R$ is
reduced, while it becomes smaller upon reducing $\Gamma_L$.
Nevertheless, as long as $\gamma_0$ is sufficiently large, one can
achieve in both cases a correlation coefficient close to unity.

\subsection{Source and drain currents}

The correlation between the source current $I_L$ and the detector
current follows by inserting \eqref{CNL} into \eqref{Cjx}.  Unless
$\Gamma_L=\Gamma_R$, the resulting expression is neither symmetric nor
anti-symmetric, so that the spectral density is complex and reads
\begin{align}
\label{CjL}
C_{jI_L}(\omega)
={}&
\nu\gamma_0\langle I\rangle \,
\frac{-i\omega+\Gamma_L-\Gamma_R}{\omega^2+(\Gamma_L+\Gamma_R)^2}
\\
={}& C_{jI_R}^*(\omega) .
\label{CjR}
\end{align}
The second equation represents the corresponding expression for the
SET drain and is obtained accordingly.  For a symmetric SET with
$\Gamma_L=\Gamma_R$, this correlation function vanishes in the limit
$\omega\to 0$ in accordance with the asymmetric behavior as function
of time, cf.\ Eqs.~\eqref{CNR} and \eqref{CNL}.  This implies that the
corresponding correlation for the long-time limit of the counting
statistics vanishes as well, $\lim_{t\to\infty}\langle n,N_{L/R}\rangle=0$.

A more detailed analysis of Eq.~\eqref{CjL}
reveals that the correlation coefficient $r_{jI_L}$ may assume the
appreciable value $\sqrt{1/2}$.  The limitation is due to the fact
that the detector is sensitive to both the source and the drain
current.  Because tunnel events at source and drain alternate, they
are not statistically independent.  Thus, there is no contradiction in
$|r_{jI_L}|$ exceeding the value $1/2$.  We depict in
Fig.~\ref{fig:correlations}(b) a quantitative analysis for the
symmetric situation.  It shows that also here the condition
$\gamma_0\geq 100\Gamma,100\omega$ must be fulfilled in order to
achieve a significant correlation.  Moreover, at frequencies below the
SET rate, $\omega\lesssim\Gamma$, both currents become statistically
independent, because the detector ``sees'' merely the average
population of the SET.

\subsection{Ramo-Shockley current}

According to the Ramo-Shockley theorem, the experimentally relevant
SET current is the one given by Eq.~\eqref{RStheorem}, i.e., $I =
\alpha I_L+\beta I_R$.  From charge conservation follows $-I_L+\dot
N+I_R$, while $C_{\dot N \dot N}(\omega) = \omega^2C_{NN}(\omega)$.
This yields for the total current the correlation function
\cite{Blanter2000a}
\begin{align}
C_{II}(\omega)
={}& \alpha C_{I_LI_L} + \beta C_{I_RI_R} -\alpha\beta\omega^2 C_{NN}
\\
={}& \frac{(1-2\alpha\beta)\omega^2+\Gamma_L^2+\Gamma_R^2}
          {\omega^2+(\Gamma_L+\Gamma_R)^2}
,
\end{align}
which we need for normalizing $C_{jI} = \alpha C_{jI_L}+\beta
C_{jI_R}$.  Writing $C_{jI_\ell}$ for $\ell=L,R$ in terms of
$C_{NI_\ell}$, see Eq.~\eqref{Cjx}, we obtain
\begin{equation}
C_{jI}(\omega) = \nu\gamma_0 \langle I\rangle \,
\frac{\Gamma_L-\Gamma_R-i\omega(\alpha-\beta)}{\omega^2+4\Gamma^2}\,.
\end{equation}
Interestingly enough, $C_{jI}$ depends on the difference of the tunnel
rates, $\Gamma_L-\Gamma_R$, and on the difference of the Ramo-Shockley
coefficients, $\alpha-\beta$.  It vanishes for a completely symmetric
setup.  Thus, we face the surprising situation that the realization of
the stochastic process $I(t)$ can be determined by
measuring $j(t)$, despite that both quantities are uncorrelated.  This
underlines that the determination of the SET current by charge
detection must rely on implicit knowledge about the transport process,
in the present case its unidirectionality.

An intuitive explanation for the lack of correlations in the symmetric
case can be derived from the fact that unidirectional transport
through a symmetric SET can be mapped to a Poisson process with half
charges \cite{Elattari2002a}.  The events of this process are
electron tunnelings from the source to the SET and from the SET to the
drain, which both contribute with a half electron to the total
current $I$.  The waiting time between two subsequent events is
exponentially distributed with equal mean time \cite{Brandes2008a}.
Thus, with any of the two tunnel events, the detector may switch in either
direction, from on to off or back.  Then for symmetry reasons the
correlation between the detector and the total SET current $I$ must
vanish.

Going beyond the symmetric situation, we find that the imaginary part
$\Im C_{jI}(\omega)$ is finite and proportional to the difference of
the Ramo-Shockley coefficients, $\alpha-\beta$.  This in principle
allows one to determine the ratio of the barrier capacitances $c_L$
and $c_R$.  Analyzing the
correlation coefficient $r_{jI}(\omega)$ for $|\alpha-\beta|\ll1$
reveals that such measurement is possible under the following
conditions.  First, as for all other SET-detector correlations, the
Fano factor $F_\text{QPC}$ must lie significantly above the Poissonian
value which requires $\gamma_0\gtrsim100\Gamma$.  Moreover, the
measurement frequency must be so large that the SET is not in
its zero-frequency limit, $\omega \gtrsim \Gamma$.  Then
$\Im r_{jI}\approx |\alpha-\beta|$ in an intermediate frequency range.
The data shown in Fig.~\ref{fig:correlations}(c) visualize this
estimate.
In recent experiments on monitoring SET currents
\cite{Gustavsson2007a, Fujisawa2006a, Fricke2007a}, the relevant
frequencies were of the order 10\,kHz.  In this regime, it is possible
to record time-resolved measurements and to subsequently obtain the
frequency dependent correlation coefficient by numerical data
processing.

\section{Discussion}

We have studied correlations between the currents of a charge
detector interacting capacitively with a SET.  As a simple fully
classical model, we have employed as detector a QPC in the
weak-tunneling limit.  Then electron transport through the detector
constitutes a Cox process, i.e., a Poisson with a stochastic rate for
which all correlation function can be obtained analytically.

The fundamental correlation upon which the measurement idea is based,
is the one between the detector and the SET occupation. It grows with
the Fano factor of the detector current, while no particular features
of the SET enter.  The need for super Poissonian detector noise
indicates the requirement for switching between large periods of
conducting and non-conducting behavior.  On the quantitative level, on
the order of 100 electrons should flow during a conducting period.
Then the current noise of the detector exhibits significant bunching.

For a symmetric SET whose tunnel barriers possess equal tunnel rates
and equal Ramo-Shockley factors, the detector current is independent
of the total SET current, i.e, the average between the source and
drain current.  This can be understood by mapping of the symmetric SET
to a Poisson process with ``half charges'', because in this picture
both source-SET tunneling and SET-drain tunneling contribute equally
to the total current.  Beyond the fully symmetric situation,
correlations emerge.  Most interesting is a contribution proportional
to the difference of the Ramo-Shockley coefficients which, thus, in
principle can be measured.
In contrast to the total current, the source current and the drain
current correlate with the detector even for a symmetric setup.  The
correlation is limited to $\sqrt{1/2}$, which still is an appreciable
value.

In conclusion, already a simple model for mesoscopic charge monitoring
exhibits interesting correlations that should be measurable readily with
present setups.  Many more may be predicted for coupled conductors that
allow for quantum features such as electrons in delocalized orbitals.

\begin{acknowledgement}
This work was supported by the Spanish Ministry of Economy and
Competitiveness through grant No.\ MAT2011-24331.
\end{acknowledgement}




\begin{thebibliography}{10}

\bibitem{Gustavsson2006a}
S. Gustavsson, R. Leturcq, B. Simovi\v{c}, R. Schleser, T. Ihn, P. Studerus, K.
  Ensslin, D.~C. Driscoll, and A.~C. Gossard, Phys. Rev. Lett. {\bf 96},
  076605  (2006).

\bibitem{Fujisawa2006a}
T. Fujisawa, T. Hayashi, R. Tomita, and Y. Hirayama, Science {\bf 312},  1634
  (2006).

\bibitem{Fricke2007a}
C. Fricke, F. Hohls, W. Wegscheider, and R.~J. Haug, Phys. Rev. B {\bf 76},
  155307  (2007).

\bibitem{Wiseman2001a}
H.~M. Wiseman, D.~W. Utami, H.~B. Sun, G.~J. Milburn, B.~E. Kane, A. Dzurak,
  and R.~G. Clark, Phys. Rev. B {\bf 63},  235308  (2001).

\bibitem{Schaller2010a}
G. Schaller, G. Kie{\ss}lich, and T. Brandes, Phys. Rev. B {\bf 82},  041303(R)
   (2010).

\bibitem{Kreisbeck2010a}
C. Kreisbeck, F.~J. Kaiser, and S. Kohler, Phys. Rev. B {\bf 81},  125404
  (2010).

\bibitem{Gurvitz1997a}
S.~A. Gurvitz, Phys. Rev. B {\bf 56},  15215  (1997).

\bibitem{Goan2001a}
H.-S. Goan, G.~J. Milburn, H.~M. Wiseman, and H.~B. Sun, Phys. Rev. B {\bf 63},
   125326  (2001).

\bibitem{Gilad2006a}
T. Gilad and S.~A. Gurvitz, Phys. Rev. Lett. {\bf 97},  116806  (2006).

\bibitem{Ashhab2009b}
S. Ashhab, J.~Q. You, and F. Nori, New J. Phys. {\bf 11},  083017  (2009).

\bibitem{Ashhab2009c}
S. Ashhab, J.~Q. You, and F. Nori, Phys. Scr. {\bf T137},  014005  (2009).

\bibitem{Sanchez2010a}
R. S\'anchez, R. L\'opez, D. S\'anchez, and M. B\"uttiker, Phys. Rev. Lett.
  {\bf 104},  076801  (2010).

\bibitem{Golubev2011a}
D.~S. Golubev, Y. Utsumi, M. Marthaler, and G. Sch\"on, Phys. Rev. B {\bf 84},
  075323  (2011).

\bibitem{Esposito2010a}
M. Esposito, U. Harbola, and S. Mukamel, Rev. Mod. Phys. {\bf 81},  1665
  (2010).

\bibitem{Campisi2011a}
M. Campisi, P. H\"anggi, and P. Talkner, Rev. Mod. Phys. {\bf 83},  771
  (2011).

\bibitem{Braggio2009a}
A. Braggio, C. Flindt, and T. Novotn\'y, J. Stat. Mech.  P01048  (2009).

\bibitem{Stark2010a}
M. Stark and S. Kohler, EPL {\bf 91},  20007  (2010).

\bibitem{Hussein2012a}
R. Hussein and S. Kohler, Phys. Rev. B {\bf 86},  115452  (2012).

\bibitem{Schaller2011a}
G. Schaller, C. Emary, G. Kiesslich, and T. Brandes, Phys. Rev. B {\bf 84},
  085418  (2011).

\bibitem{Flindt2009a}
C. Flindt, C. Fricke, F. Hohls, T. Novotn\'y, K. Neto{\v c}n\'y, T. Brandes,
  and R.~J. Haug, Proc. Natl. Acad. Sci. USA {\bf 106},  10116  (2009).

\bibitem{Ubbelohde2012a}
N. Ubbelohde, C. Fricke, C. Flindt, F. Hohls, and R.~J. Haug, Nature Comm. {\bf
  3},  612  (2012).

\bibitem{Gurvitz1996a}
S.~A. Gurvitz and {Y}a. S.~Prager, Phys. Rev. B {\bf 53},  15932  (1996).

\bibitem{Shockley1938a}
W. Shockley, J. Appl. Phys. {\bf 9},  635  (1938).

\bibitem{Ramo1939a}
S. Ramo, Proc. I. R. E. {\bf 27},  584  (1939).

\bibitem{Blanter2000a}
{Y}a. M.~Blanter and M. B\"uttiker, Phys. Rep. {\bf 336},  1  (2000).

\bibitem{Mozyrsky2002a}
D. Mozyrsky, S. Kogan, V.~N. Gorshkov, and G.~P. Berman, Phys. Rev. B {\bf 65},
   245213  (2002).

\bibitem{Ingold1992a}
G.-L. Ingold and {Y}u. V.~Nazarov,  in {\em Single Charge Tunneling}, Vol.~294
  of {\em NATO ASI Series B} (Plenum, New York, 1992), pp.\ 21--107.

\bibitem{Emary2007a}
C. Emary, D. Marcos, R. Aguado, and T. Brandes, Phys. Rev. B {\bf 76},  161404
  (2007).

\bibitem{Brandes2008a}
T. Brandes, Ann. Phys. (Leipzig) {\bf 17},  477  (2008).

\bibitem{Bagrets2003a}
D.~A. Bagrets and {Y}u. V.~Nazarov, Phys. Rev. B {\bf 67},  085316  (2003).

\bibitem{Korotkov1994a}
A.~N. Korotkov, Phys. Rev. B {\bf 49},  10381  (1994).

\bibitem{Hanke1995a}
U. Hanke, Y. Galperin, K.~A. Chao, M. Gisself\"alt, M. Jonson, and R.~I.
  Shekhter, Phys. Rev. B {\bf 51},  9084  (1995).

\bibitem{vanKampen1992a}
N.~G. van Kampen, {\em Stochastic processes in physics and chemistry}
  (North-Holland, Amsterdam, 1992).

\bibitem{Bouzas2006a}
P.~R. Bouzas, M.~J. Valderrama, and A.~M. Aguilera, Appl. Math. Modell. {\bf
  30},  1021  (2006).

\bibitem{Elattari2002a}
B. Elattari and S.~A. Gurvitz, Phys. Lett. A {\bf 292},  289  (2002).

\bibitem{Gustavsson2007a}
S. Gustavsson, M. Studer, R. Leturcq, T. Ihn, K. Ensslin, D.~C. Driscoll, and
  A.~C. Gossard, Phys. Rev. Lett. {\bf 99},  206804  (2007).

\end{thebibliography}
\end{document}